\documentclass[twocolumn]{aastex63}

\hypersetup{linkcolor=red,citecolor=blue,filecolor=cyan,urlcolor=blue}
\usepackage{graphicx,color}
\usepackage{amssymb}
\usepackage{amsmath}
\usepackage{url}
\usepackage{natbib}
\usepackage{txfonts}
\usepackage{multirow}
\usepackage{array}
\usepackage{rotating}
\usepackage{sidecap}
\usepackage{hyperref}
\usepackage{epstopdf}
\usepackage{footnote}
\usepackage{tabularx}
\usepackage{booktabs}
\usepackage{longtable}
\usepackage{tabu}
\usepackage{longtable}

\newcommand{\degree}{\ensuremath{^\circ}}

\shortauthors{Moore et al.}

\shorttitle{On Making Magnetic-Flux-Rope $\Omega$  Loops }
\shortauthors{Moore et al.}

\begin{document}
	\title{On Making Magnetic-Flux-Rope $\Omega$ Loops for Solar Bipolar Magnetic Regions of All Sizes by
		Convection Cells
	}

\correspondingauthor{Sanjiv K. Tiwari}
\email{tiwari@lmsal.com}

\author[0000-0002-5691-6152]{Ronald L. Moore}
\affiliation{Center for Space Plasma and Aeronomic Research (CSPAR), UAH, Huntsville, AL 35805, USA}
\affiliation{NASA Marshall Space Flight Center, Huntsville, AL 35812, USA; [Email: ronald.l.moore@nasa.gov]}

\author[0000-0001-7817-2978]{Sanjiv K. Tiwari}
\affil{Lockheed Martin Solar and Astrophysics Laboratory, 3251 Hanover Street, Bldg. 252, Palo Alto, CA 94304, USA}
\affil{Bay Area Environmental Research Institute, NASA Research Park, Moffett Field, CA 94035, USA}

\author[0000-0001-7620-362X]{Navdeep K. Panesar}
\affil{Lockheed Martin Solar and Astrophysics Laboratory, 3251 Hanover Street, Bldg. 252, Palo Alto, CA 94304, USA}
\affil{Bay Area Environmental Research Institute, NASA Research Park, Moffett Field, CA 94035, USA}

\author[0000-0003-1281-897X]{Alphonse C. Sterling}
\affiliation{NASA Marshall Space Flight Center, Huntsville, AL 35812, USA}

\begin{abstract}

We propose that the flux-rope $\Omega$ loop that emerges to become any bipolar magnetic region (BMR) is made by a convection cell of the $\Omega$-loop's size from initially-horizontal magnetic field ingested through the cell's bottom.  This idea is based on (1) observed characteristics of BMRs of all spans ($\sim$ 1000 km to $\sim$ 200,000 km), (2) a well-known simulation of the production of a BMR by a supergranule-size convection cell from horizontal field placed at cell bottom, and (3) a well-known convection-zone simulation.  From the observations and simulations, we (1) infer that the strength of the field ingested by the biggest convection cells (giant cells) to make the biggest BMR $\Omega$ loops is $\sim$ 10$^3$ G, (2) plausibly explain why the span and flux of the biggest observed BMRs are $\sim$ 200,000 km and $\sim$ 10$^{22}$ Mx, (3) suggest how giant cells might also make ``failed-BMR" $\Omega$ loops that populate the upper convection zone with horizontal field, from which smaller convection cells make BMR $\Omega$ loops of their size, (4) suggest why sunspots observed in a sunspot cycle's declining phase tend to violate the hemispheric helicity rule, and (5) support a previously-proposed amended Babcock scenario for the sunspot cycle's dynamo process.  Because the proposed convection-based heuristic model for making a sunspot-BMR $\Omega$ loop avoids having $\sim$ 10$^5$ G field in the initial flux rope at the bottom of the convection zone, it is an appealing alternative to the present magnetic-buoyancy-based standard scenario and warrants  testing by high-enough-resolution giant-cell magnetoconvection simulations.
\end{abstract}


\section{Introduction} \label{sec:intro}

All of the magnetic field on the Sun evidently comes from magnetic-flux-rope $\Omega$ loops that bubble up from below the surface, i.e., from below the photosphere \citep{zwaa87,fan09,ishi10,van15}.  The magnetic field in and above the photosphere continually evolves via emergence of new $\Omega$-loop field, movement of the field's feet by flows in and below the photosphere, and submergence of extant field \citep[e.g.,][]{moor85}.  The evolving magnetic field modulates the Sun's luminosity, results in the mega-Kelvin hot corona and its solar-wind outflow, and explodes to make coronal mass ejections, flares, and myriad smaller explosions that continually erupt all over the Sun \citep{with77,vaia78,shib07,raou10,moor11,inne13,tiw19,pane19,ster20}.  Thus, (1) the dynamo process that generates the pre-emergent magnetic field and (2) the process that forms the pre-emergent field into flux-rope $\Omega$ loops together fuel all solar magnetic activity and consequent space weather and space climate.  For this reason, and because both processes are far from being entirely understood \citep{fan09,spru11,char20}, pinning down how these two processes work continues to be a major endeavour of solar astrophysics.  The considerations of the present paper concern both processes.

The present paper considers certain well-established observed characteristics of solar bipolar magnetic regions (BMRs) of all sizes in combination with two well-known numerical simulations of the free convection in the convection zone, one with magnetic field and the other with no magnetic field.  The simulation with magnetic field is of the production of a BMR by the emergence of an $\Omega$ loop made by a convection cell of the diameter of a small supergranule, $\sim$ 20,000 km across.  The convection cell makes the $\Omega$ loop by ingesting horizontal field placed at the bottom of the cell, 20,000 km below the photosphere top of the cell.  The simulation with no magnetic field is of the global convection zone, from the bottom of the convection zone at 0.7 R$_{Sun}$ up to 15,000 km below the photosphere.  From the considered observed characteristics of BMRs together with the simulations, we propose that the flux-rope $\Omega$ loop that emerges to become a BMR of any size -- from the littlest to the biggest -- is made by a convection cell of that size.  With schematic drawings (cartoons) that are suggested by the observations and simulations, we envision how the flow in a BMR-making convection cell of any size bends and twists initially-horizontal ingested field into a flux-rope $\Omega$ loop as the cell's central upflow carries the top of the loop up to the surface.

We also propose how the cell-bottom horizontal field -- from which a convection cell of any size makes a BMR flux-rope $\Omega$ loop of its size -- comes to be present at the bottom of the cell.  For this, we employ the convection zone's so-called giant convection cells.  A giant cell is any convection cell that is big enough to span the entire 200,000 km vertical extent of the convection zone \citep{simo68,hath13}.  The giant cells in the global simulation of the convection zone range in width from $\sim$ 100,000 km to $\sim$ 200,000 km.  

From the cartoons and the global simulation of the convection zone, we reason that only a few of the giant cells that sit on either the northern or the southern global-dynamo-generated toroidal field band at the bottom of the convection zone may ingest a stitch of toroidal field from which it makes a twisted-field flux-rope $\Omega$ loop that spans the giant cell and emerges to become a BMR of that size.  Observed recently-emerged BMRs of that size are the Sun's biggest single-bipole sunspot active regions.  With similar cartoons, we envision how most giant cells that sit on a toroidal field band and ingest a stitch of toroidal field might make from it a cell-spanning $\Omega$ loop that -- instead of being a twisted flux rope -- is a loose bundle of many separate flux tubes that do not immediately emerge as BMRs.  We conjecture that the tops of these failed-BMR $\Omega$ loops reside in the upper half of the convection zone for about a month, waiting for convection downflows to pump them down to the bottom of the convection zone.  During their stay in the upper half, they are the horizontal field that convection cells of all sizes smaller than giant cells feed on to make flux-rope $\Omega$ loops of their size that emerge as BMRs of that size.

The envisioned way that a giant cell makes an $\Omega$ loop of its size is commensurate with the flows in giant cells in the simulated global convection zone.  We also point out how both the envisioned way of producing the biggest BMR flux-rope $\Omega$ loops and the global convection-zone simulation support an amended Babcock solar dynamo scenario.

 \section{BMR Observations}\label{obs}  

New (recently-emerged) BMRs of all sizes have grossly similar magnetic form, each being an arch of magnetic field rooted in the BMR's two domains of opposite-polarity magnetic flux.  One measure of the size of a BMR is the BMR's span D given by the distance between the centroids of the two domains of opposite polarity.  By this measure, new BMRs range in size (span D) in a continuous spectrum across about two orders of magnitude \citep{van15}.  The littlest new BMRs are the size of a granule (D $\sim$ 10$^3$ km) \citep{ishi10}; the biggest recently-emerged BMRs are the biggest first-disk-passage single-bipole sunspot active regions (D $\sim$ 2 x 10$^5$ km) \citep{wang89}.

\begin{figure*}
	\centering
	\includegraphics[trim=0.5cm 4.7cm 0.1cm 2cm,clip,width=\linewidth]{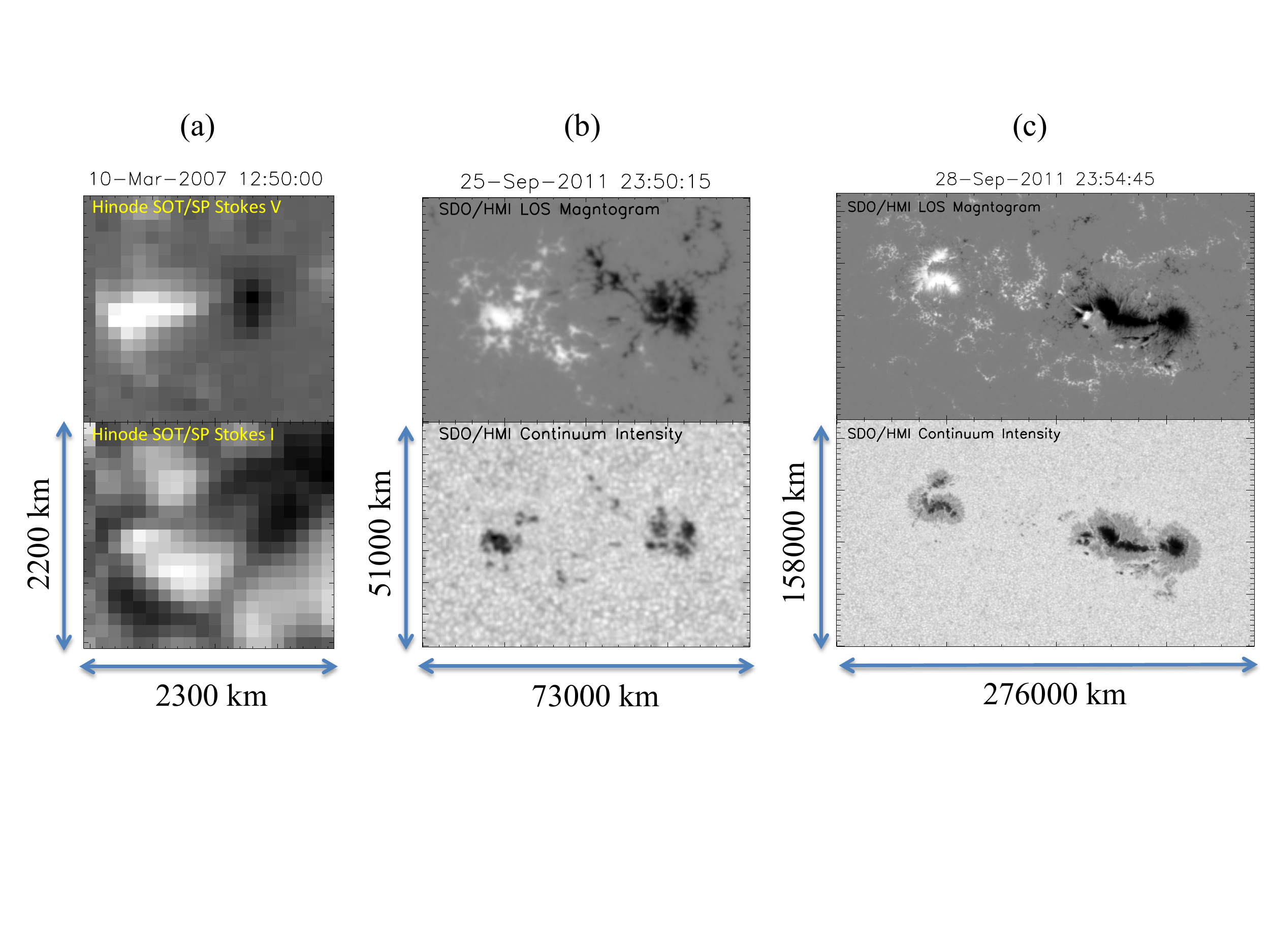}
	\caption{Examples of BMRs of three different sizes near their times of maximum magnetic flux.  (a) Stokes-V (upper panel) and Stokes-I (lower panel) of a granule-size BMR scanned by Hinode SOT/SP.  (b) LOS magnetogram (upper panel) and continuum intensity image (lower panel) of a supergranule-size BMR observed by SDO/HMI.  (c) Same as (b), but for a giant-cell-size BMR.  Each BMR is in the northern hemisphere, and is within 20$\degree$ from solar disk center.  East-west and north-south spans of the fields of view are given in the lower panels.} \label{fig1}
\end{figure*}

    A BMR's magnetic arch has the overall shape of an elongated dome that covers a roughly elliptical area typically having its major diameter about twice its minor diameter.  That overall shape of the magnetic field holds for isolated recently-emerged BMR's big enough to have sunspots, for new emerged BMRs too little to have sunspots (i.e., the BMRs called ephemeral regions), and for BMRs even smaller than ephemeral regions.  In Figure \ref{fig1} we show typical BMRs of three different sizes -- of the size of a granule, of the size of a supergranule, and of the size of a giant cell.  For more examples of typical single-bipole sunspot-active-region BMRs see Figures 1 and 2 of \cite{fan09} and Figure 1 of \cite{van15}.  For examples of ephemeral-region BMRs see Figure 7.4 of \cite{mart77}.  As Figure \ref{fig1} illustrates, granule-size BMRs are like all bigger BMRs in that the overall three-dimensional magnetic field evidently has the same elongated elliptical dome shape  \citep{ishi10}.
     
As a BMR of any size emerges, the opposite-polarity flux centroids are closest together at the start of emergence, continually move apart as more field emerges, and stop moving apart at the end of emergence \citep{ishi10,van15}.  The similar spreading apart of the opposite-polarity centroids and the similar shape of the emerged bipolar field of BMRs of all sizes are the basic evidence that BMRs of all sizes are similar magnetic structures that all emerge the same way, each resulting from the emergence of a flux-rope $\Omega$ loop from below the photosphere \citep{van15}.

\subsection{Granule-Size BMRs}

          Photospheric vector magnetograms from the Solar Optical Telescope (SOT) Spectropolarimeter (SP) on Hinode \citep{tsun08,ichi08,suem08,kosu07} revealed that in quiet regions and coronal holes, there is a granule-size BMR in about one out of every ten granules \citep{lites08,ishi08}.  Averaged over the BMR's granule-size area, the strength of the magnetic field in a granule-size BMR is $\sim$ 100 G  \citep{lites08}.  So, the total unsigned magnetic flux in a granule-size BMR is $\sim$ 10$^{18}$ Mx ($\sim$ BD$^2$, where B is the field strength, $\sim$ 100 G, and D is the distance between the opposite-polarity flux centroids, $\sim$ 10$^8$ cm).  The magnetic flux $\Phi$ in the $\Omega$ loop that emerges to become a BMR is half the BMR's total unsigned flux, $\sim$ 5 $\times$ 10$^{17}$ Mx for the $\Omega$ loops that emerge to become granule-size BMRs.  A typical granule-size BMR has about the span of a granule ($\sim$ 10$^8$ cm) and lasts about as long as a granule ($\sim$ 300 s) \citep{titl89,lites08,ishit10}.  
          
\cite{ishi10} tracked the emergence of a granule-size BMR in SOT/SP vector magnetograms and intensity images.  The BMR emerged in an emerging new granule convection cell.  The BMR was at first detected as an elongated patch of only horizontal magnetic field, about 600 km long and 300 km wide, with field strength of a few hundred Gauss.  In about 2 minutes, as the granule grew, the BMR grew to about 1,300 km long and 700 km wide, and at each end of the BMR the field had an obvious vertical component of the polarity required for an emerging $\Omega$ loop in which the field had the observed horizontal direction.  At that time, the BMR nearly spanned the granule and each end was in or near a dark intergranular lane at an edge of the granule, and so was in or being swept into the downflow at that edge of the granule convection cell.  Both the BMR's being in the granule and growing in synchrony with the granule to span the granule and the observed growth and evolving form of the BMR's vector magnetic field give the impression that the BMR is made by an emerging flux-rope $\Omega$ loop that is formed in and by the plasma flow in the granule convection cell.

Because BMRs of all sizes have similar magnetic form and similar progression of emergence, the apparent production of the granule-size BMR by the granule convection cell suggests that each bigger flux-rope $\Omega$ loop that emerges to make a correspondingly bigger BMR is made in the same way as a granule-size BMR by a convection cell of the size of the BMR.

\subsection{Biggest Recently-Emerged BMRs}

     In each 11-year sunspot cycle, nearly all BMRs big enough to have sunspots (leading-polarity flux $\Phi$ $>\sim 3 \times 10^{20}$ Mx, \citep{wang89} occur at latitudes less than 30\degree north and south \citep{hath15lrsp}.  In nearly all sunspot BMRs -- and hence in the flux-rope $\Omega$ loops that emerge to make them -- the magnetic field is directed roughly east-west.  That is, the flux of one polarity leads the flux of the other polarity in the direction of solar rotation.  During each cycle, nearly all of the sunspot BMRs in the northern hemisphere have leading flux of one polarity and nearly all in the southern hemisphere have leading flux of the opposite polarity.  During the next 11-year sunspot cycle, the east-west direction of the field in the BMR emerged $\Omega$ loops in each hemisphere is the reverse of what it was in the previous cycle.  These solar-cycle hemispheric rules for the magnetic field in sunspot-BMR $\Omega$ loops constitute what is called Hale's law \citep{haln25}; see the review by \cite{van15}.
     
Hale's law is generally thought to show that during each sunspot cycle the sunspot-BMR $\Omega$ loops in each hemisphere bubble up to the surface from a global band of toroidal magnetic field below that has the east-west direction of the field in the $\Omega$ loops and that is somehow generated by a global dynamo process \citep{fan09,van15}.  It is now widely thought that the toroidal field band sits at the bottom of the convection zone and builds up there until strands of it become buoyant enough to overcome some downward-pushing restraint and bubble up through the convection zone \citep[e.g.,][]{parker55,parker75}; see the review by \citep{fan09}.  One idea for the downward-pushing restraint is that it might be the dynamic pressure [$\rho(v_c)^2$/2, where $\rho$ is the mass density and v$_c$ is the flow speed] of the convection-cell downflows at the bottom of the convection zone \citep{fan09}.  To the contrary, from MHD simulations of the buoyant rise of a flux rope through the convection zone and comparison of aspects of the resulting simulated rising flux-rope $\Omega$ loop with corresponding aspects of observed sunspot BMRs, it has been inferred that the strength of the field in the initial flux rope at the base of the convection zone needs to be $\sim$ 10$^5$ G, giving the initial flux rope a magnetic pressure and consequent buoyancy far greater than the dynamic pressure of the convection downflows expected at the bottom of the convection zone \citep{fan09}.

From each of 2700 recently-emerged sunspot BMRs in full-disk magnetograms from the National Solar Observatory/Kitt Peak, \cite{wang89} measured the flux $\Phi$ of the leading-polarity flux domain and the separation distance D of the centroid of that domain from the centroid of the opposite-polarity trailing flux domain.  The (Log D, Log $\Phi$) scatter plot of the measured values shows that, on average, the greater a BMR's centroid separation, the greater its flux.  From this scatter plot, the linear least squares fit for Log $\Phi$ as a function of Log D gives $\Phi$ = 4 $\times 10^{20} D^{1.3}$ Mx, where the length unit for D is the length spanned by 1 heliocentric degree on the solar surface (1$\degree$ = 12,150 km).  In this plot, D ranges from about 10,000 km to about 200,000 km, and $\Phi$ ranges from 3 $\times 10^{20}$ Mx to about 3 $\times 10^{22}$ Mx.  The solar convection zone is about 200,000 km deep \citep[e.g.,][]{fan09}.  So, the BMR measurements of \cite{wang89} show that in the biggest recently-emerged BMRs, the separation distance of the opposite-polarity flux centroids roughly equals the depth of the convection zone.  This observation, with the notion that the flux-rope $\Omega$ loop that emerges to make a BMR of centroid-separation distance D is made by a convection cell of diameter D, suggests that the convection zone's biggest convection cells are about as wide as the convection zone is deep, and that the two flux centroids in the biggest recently-emerged BMRs, i.e., the two legs of the emerged $\Omega$ loop, have been swept into, and are centered on, downflows on opposite sides of one of the biggest cells of solar convection.  \cite{moor00} pointed out that big BMRs might have this connection with giant cells, the convection zone's biggest convection cells, and, if so, the biggest giant cells would be about as wide as the convection zone is deep.

 \section{Two Simulations}

\subsection{A Simulation of the Production of a BMR by a Convection Cell}

    \cite{stei12} simulated the production of a BMR by the emergence of an $\Omega$ loop made by a convection cell of the diameter of a small supergranule ($\sim$ 20,000 km across).  They used a simulation of magnetoconvection in the top 20,000 km of the convection zone, computed in a square box 48,000 km wide and 20,000 km deep.  The top of the box mimics the top of the photosphere.  At the start of the simulation, there was no magnetic field in the box, only steady free convection.  Uniform horizontal field of gradually increasing strength, up to 10$^3$ G, was introduced at the bottom.  The horizontal field was skewed 30\degree\ from east-west.  In about a day, a $\sim$ 20,000 km wide convection cell, centered near the center of the box and spanning the 20,000 km depth of the box, had ingested a stitch of the initially horizontal field, and had bent and twisted it into an $\Omega$ loop that had each leg in a downflow on opposite sides of the cell, and the top of the $\Omega$ loop was near the photosphere.  In three more days, the $\Omega$ loop completed its emergence to make a BMR having its opposite-polarity fluxes concentrated in downflows $\sim$ 20,000 km apart, having about the 30\degree\ tilt of the initial horizontal field at the bottom, and having sunspot pores in flux concentrations of either polarity.  

     A sunspot pore is a sunspot that is too little to have penumbra \citep{bruz77}.  The smallest sunspot BMRs are of the size of the simulated BMR, $\sim$ half the size of the BMR in Figure \ref{fig1}b, and their sunspots are typically pores.
     
The observed granule-size BMR analyzed by \cite{ishi10} is similar in magnetic form and progression  of emergence to observed BMRs of all sizes, and is observed to have its opposite-polarity ends in downflows on opposite sides of the granule convection cell in which it emerges.  The simulated BMR produced in the \cite{stei12} simulation is also similar in magnetic form and progression of emergence to observed BMRs of all sizes, and has its opposite-polarity ends in downflows on opposite sides of the small-supergranule-size convection cell in which it emerges.  In that way, observed BMRs and the \cite{stei12} simulation together suggest that the flux-rope $\Omega$ loop for a BMR of any size -- from the size of a granule to the size of a giant cell -- is made in the same way as in the simulation.  That is, the BMR observations and the BMR-production simulation together suggest that, for BMRs of all sizes, a convection cell of the size of the BMR makes the $\Omega$ loop by ingesting horizontal field that happens to be present at the bottom of the cell.

\subsection{A Global Simulation of the Convection Zone}

   \cite{mies08} present a computer-generated MHD simulation of all but the top 15,000 km of the global solar convection zone.  The simulation extends from the bottom of the convection zone at 0.7 R$_{Sun}$ to 15,000 km below the photosphere.  The simulated convection zone is full of continually evolving big convection cells that at the top have diameters ranging from $\sim$ 200,000 km down to $\sim$ 10 times smaller.  So, the simulation's biggest giant cells are, as anticipated by \cite{moor00}, about as wide as the convection zone is deep.  They are of the size expected if the biggest-BMR emerged $\Omega$ loops that have been observed had each leg trapped in a downflow on opposite sides of one of the convection zone's biggest giant cells.

The stronger of the simulation's downflows are at edges of giant cells, are $\sim$ 200,000 km apart, reach to the bottom of the convection zone, and persist for about a month or more.  The speed v$_c$ of these giant-cell downflows at $\sim$ 200,000 km below the photosphere, i.e., near the bottom of the simulated convection zone, is $\sim$ 10$^3$ cm s$^{-1}$.  The plasma mass density $\rho$ at this depth is about 0.2 gm cm$^{-3}$ \citep{guen92}.  So, the downward dynamic pressure $\rho(v_c)^2/2$ of the giant-cell downflows at the bottom of the simulated convection zone is $\sim$ 10$^5$ dyne cm$^{-2}$.  Hence, an upper bound on the strength B$_{bcz}$ of the horizontal field that might be confined (held down) to the bottom of the convection zone by the simulation's giant-cell downflows is the field strength at which the magnetic pressure B$^2/8\pi$ equals the dynamic pressure of the simulation's giant-cell downflows at the bottom \citep{fan09}, which gives B$_{bcz} < 2 \times 10^3$ G.

This field-strength-limit value given by the convection-zone simulation is an upper bound on the allowed field strength of the global-dynamo-generated toroidal field band at the bottom of the convection zone.  That the convection-zone simulation sets this limit at 2 $\times 10^3$ G suggests the following four things.  First, the 2 $\times  10^3$ G limit suggests that the magnetic field that becomes a biggest-BMR flux-rope $\Omega$ loop does not start out as a $\sim 10^5$ G toroidal flux rope that was somehow made in the toroidal field band, as is often assumed in simulations of the production of big-BMR flux-rope $\Omega$ loops \citep{fan09}.  Second, the 2 $\times 10^3$ G limit suggests that the field to be made into a biggest-BMR flux-rope $\Omega$ loop instead comes directly from the toroidal field band.  Third, the 2 $\times 10^3$ G limit, by suggesting that the field that becomes such an $\Omega$ loop comes directly from the toroidal field band, also suggests that the embryo $\Omega$-loop field initially has close to the same strength as the field in the toroidal field band ($\sim 10^3$ G, $<< 10^5$ G).  Fourth, the 2 $\times 10^3$ G limit suggests that the toroidal field band becomes susceptible to having a biggest-BMR $\Omega$ loop made from some of it by one of the convection zone's biggest giant convection cells when the strength of the toroidal-band field has built up to  $\sim 10^3$ G.

\section{Envisioned Flux-Rope $\Omega$-Loop Production}

     Based on the above points from observations of BMRs, the \cite{stei12} simulation of the production of a BMR by a convection cell, and the \cite{mies08} global convection-zone simulation, in this Section we present in detail our proposed way of making the $\Omega$ loops for the biggest BMRs (Section 4.1) and the $\Omega$ loops for all smaller BMRs ranging down to the size of those observed in granules (Section 4.2).

\subsection{Biggest Flux-Rope $\Omega$ Loops}

\begin{figure*}
	\centering
	\includegraphics[width=0.7\linewidth]{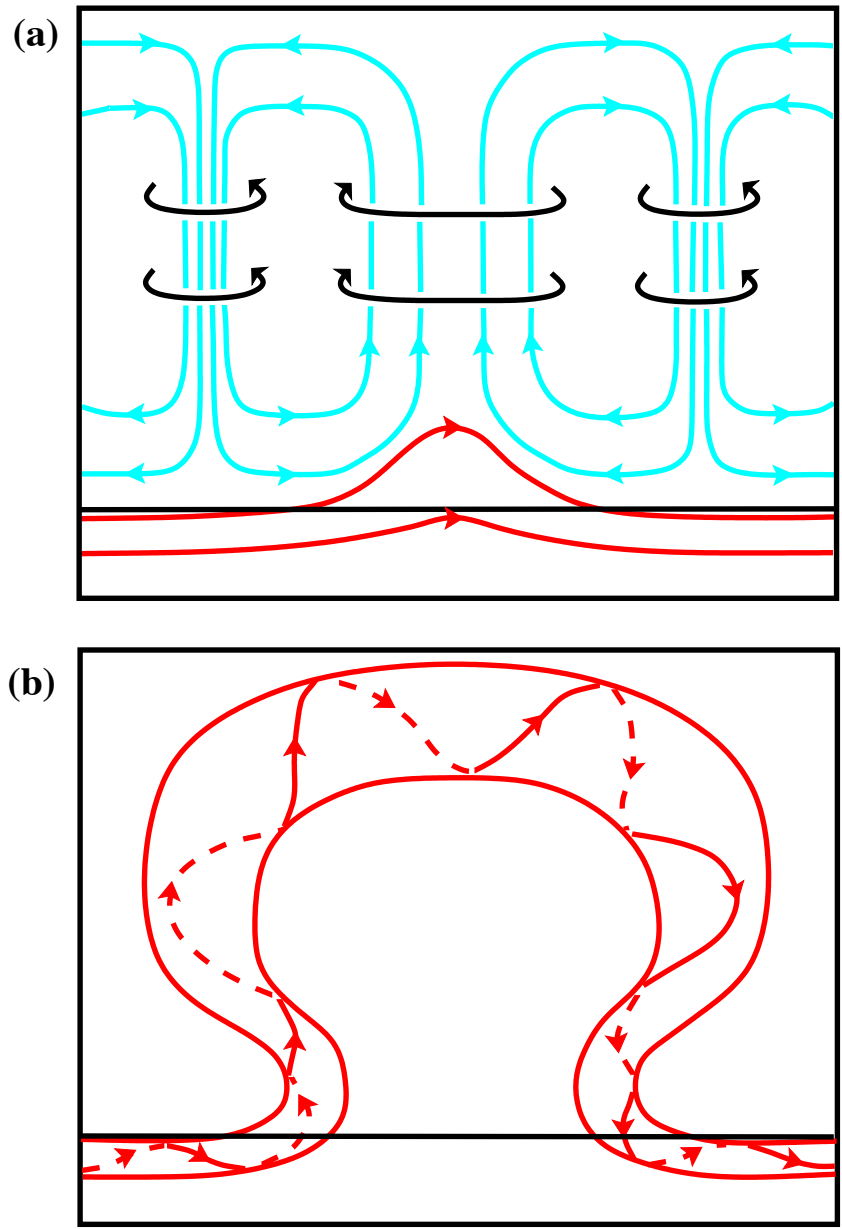}
	\caption{Schematic depiction of the production of a biggest-BMR flux-rope $\Omega$ loop by a giant cell from ingested toroidal magnetic field.  The drawings are drawn to scale for a giant cell that is 200,000 km wide, spans the full 200,000 km depth of the convection zone, sits $\sim$ 20$\degree$ north of the equator, and is viewed from the south.  The drawings depict convection flows and magnetic field in or centered on a vertical east-west cross-section.  The top of each panel is the photosphere.  The bottom of each panel is 40,000 km below the convection zone's bottom, which is marked by the horizontal black line low in the panel. [The black line is supposed to mark the middle of the tachocline transition from the convection zone to the stable radiative interior.  The tachocline is $\sim$ 30,000 km thick  \citep{thom03}.]  Blue curves are convection-flow streamlines in the plane of the cross-section.  Black curves are convection-flow streamlines in horizontal planes orthogonal to the plane of the cross-section.  Red curves are magnetic field lines that in (a) are in the plane of the cross-section and in (b) are on the surface of the flux-rope $\Omega$ loop.  In (a), at half a month after the birth of the giant cell, a flux-tube stitch of west-pointing toroidal field has welled up into the bottom of the giant cell and has no twist yet.  In (b), half a month later, the produced $\Omega$ loop's upper part has been given left-handed twist, and the same amount of right-handed twist has been given to the $\Omega$ loop's lower legs and their extensions into the global toroidal field band.} \label{fig2}
\end{figure*}

     The biggest recently-emerged BMRs have opposite-polarity-flux centroid separation distance D $\sim$ 200,000 km, and have leading-polarity magnetic flux $\Phi$ $\sim 10^{22}$ Mx \citep{wang89}.  Figure \ref{fig2} schematically depicts our scenario for how the flux-rope $\Omega$ loop that rises up through the photosphere to become such a BMR might be made by the flows in a giant convection cell from an upward protrusion in the band of toroidal field that has been built up at the bottom of the convection zone by the global dynamo process.  Figure \ref{fig2}a depicts the envisioned situation about half a month after the birth of the giant cell that sits in the center of the drawing.  Because there is no downflow but only upflow in the middle of the giant cell, the buoyancy of the toroidal field band is less restrained under the middle of the giant cell than under the downflows on the sides.  Because the toroidal field band's buoyancy was less restrained there, a patch of the toroidal field band buoyed up into the middle of the bottom of the giant cell, as depicted in Figure \ref{fig2}a.  In drawing the upward protrusion of the field band in Figure \ref{fig2}a, we assumed that before the birth of this giant cell there had been downflow over where the middle of this giant cell now sits that had kept the now-protruding patch of the field band from protruding, that is, kept it pushed to the bottom of the convection zone.  We assumed that, starting from the birth of the giant cell, that patch gradually bulged up, moving at of order the Alfv\'en speed in it and/or at of order the speed of the convection flow within $\sim$ 20,000 km of the bottom of the simulated convection zone, both of which are $\sim$ 10 m s$^{-1}$ \citep[][and for toroidal field strength $\sim 2 \times 10^3$ G]{mies08}.
     
Figure \ref{fig2}a sketches the envisioned flow field and magnetic field in or centered on a vertical east-west cross-section that cuts through the middle of the central giant convection cell and part way through an adjacent giant cell on each side that contributes to the concentrated downflow on its side of the central giant cell.  The vertical east-west cross-section sits at $\sim 20\degree$ latitude in the northern hemisphere and is viewed from the south: east is to the left and west is to the right.  The curvature of spherical surfaces of constant heliocentric radius is ignored: the top and bottom of the convection zone are drawn as straight lines.  The photosphere is the top edge of the panel, and the bottom of the convection zone is the horizontal black line near the bottom of the panel (this line is placed in the middle of the tachocline: see Figure \ref{fig2} caption and Section 4.2).  The drawing is to scale, spanning 240,000 km in height and 300,000 km east-west.  The bottom of the convection zone is placed 200,000 km below the photosphere, and the central giant cell is drawn 200,000 km wide and $\sim$ 200,000 km tall, spanning nearly the entire vertical extent of the convection zone, to be representative of the biggest giant convection cells in the convection-zone simulation of \cite{mies08}.

     The two red curves near the bottom of Figure \ref{fig2}a outline the two-dimensional profile of the toroidal field band and its upward protrusion traced in the vertical east-west cross-section.  The plane of the cross-section is the plane of north-south symmetry of the protrusion: the height of the protrusion symmetrically decreases north and south with distance from that vertical plane.  Under the two giant-cell downflows, outside the protrusion, the toroidal field band is 15,000 km thick vertically, and is supposed to be wider north-south than the crudely circular giant cell, which is 200,000 km in diameter.  The middle of the toroidal-field protrusion is 30,000 km thick vertically, and is reasonably $\sim$ 30,000 km wide horizontally north-south.  That is, where the protruding flux tube has its maximum protrusion, at the center of the bottom of the giant cell, the flux tube's diameter d is $\sim$ 30,000 km.  There, the flux tube's area (normal to the field) ($\sim$ d$^2$) is reasonably a few times larger than that flux tube's area under the downflows.  If the strength of the toroidal field is $\sim 2 \times 10^3$ G under the downflows, then, at the center of the bottom of the giant cell the field strength in the  protruding flux tube is a few times weaker, $\sim 10^3$ G.  For this field strength, the magnetic flux ($\sim Bd^2$) in the protruding flux tube is of the order of the flux in the flux-rope $\Omega$ loops that become the biggest recently-emerged BMRs, $\sim 10^{22}$ Mx.

In Figure \ref{fig2}a, the blue contours are streamlines of the giant-convection-cell plasma flow in the plane of the vertical east-west cross-section through the middle of the central giant cell.  They show that the direction of the flow is upward in the middle of the cell and downward at the edges, and that the upflow is slower and wider than the downflows.  The black contours are streamlines of horizontally swirling flows that are centered on the axis of symmetry of the central giant cell's central upflow or on the axis of symmetry of the downflow at either edge of the central giant cell.  Each swirl is the Coriolis-effect circulation resulting from the conservation of angular momentum in a rotating fluid body such as the Sun.  In the northern hemisphere viewed from above, updrafts swirl clockwise and downdrafts swirl counterclockwise.  That is, the updrafts and downdrafts swirl in the directions shown by the arrowheads on the black swirl streamlines in Figure \ref{fig2}a.

In our scenario for the $\Omega$-loop's production, we suppose that the further rise of the flux tube in Figure \ref{fig2}a and its transformation into a flux-rope $\Omega$ loop results primarily from the giant cell's convection flow and only secondarily from the flux rope's magnetic buoyancy.  After the stage depicted in Figure \ref{fig2}a, we ignore magnetic buoyancy effects, even though buoyancy might well add significantly to the rise speed of the $\Omega$ loop's top resulting from the giant cell's central updraft alone.  If the downward dynamic pressure of the giant-cell downflows is what sets the magnetic pressure (B$^2/8\pi$) of the toroidal field, perhaps the buoyancy of the protruding stitch of toroidal field in Figure \ref{fig2}a is of the same order as the upwelling convection's upward force on the stitch.  Our basic assumption is that the convection forces on the rising stitch are at least comparable to the buoyancy force, and that as a result, the kinematics of the giant-cell convection that we have envisioned in Figure \ref{fig2}a, based on the \cite{mies08} simulation of giant cells, give a useful gross picture (Figure \ref{fig2}b) of the $\Omega$ loop's evolving shape and extent in the giant cell.  That is a bold assumption.  It needs testing by appropriate numerical simulations of $\Omega$-loop production by giant cells, simulations similar to the simulation of \cite{stei12} of the production of a BMR $\Omega$ loop by a supergranule-size convection cell, but having higher spatial resolution.  Numerical simulations capable of realistically determining the importance of buoyancy relative to convection in the rise of a giant-cell $\Omega$ loop will require higher spatial resolution than present convection-zone numerical simulations have, because the thinnest flux-tube substrands of the roughly horizontal top of the rising $\Omega$ loop should be the most buoyant \citep[e.g.,][]{gilm18}.

Figure \ref{fig2}b shows the flux-rope $\Omega$ loop that the central giant cell's convection flow has made from the flux tube that is protruding from the toroidal field band in Figure \ref{fig2}a by entraining that protrusion for about half a month from the time of Figure \ref{fig2}a.  In Figure \ref{fig2}b, the middle of the protrusion has been carried up along the center line of the giant cell by the upflow to become the top of the flux-rope $\Omega$ loop, and is now about 150,000 km higher than in Figure \ref{fig2}a.  At mid-levels of the convection zone, the speed of the giant-cell upflows in the \cite{mies08} simulation is $\sim$ 100 m s$^{-1}$.  At this speed the middle of the flux tube in Figure \ref{fig2}a would be conveyed upward 150,000 km in $\sim$ 15 days.

In Figure \ref{fig2}a, the direction of the embryo-$\Omega$-loop flux tube is straight east-west.  As the upper part of the flux-tube $\Omega$ loop is carried up by the giant cell's upflow, the clockwise rotation of the flow should tilt the horizontal direction of the top part of the $\Omega$ loop away from east-west, so that the western leg of the $\Omega$ loop sits closer to the equator than the eastern leg.  If it does, then when the flux-rope $\Omega$ loop emerges through the photosphere to make a biggest recently-emerged BMR, the BMR will have a tilt from east-west having the sense of the average sunspot BMR tilt observed in the northern hemisphere, that is, it will have Joy's-law sense of tilt \citep{van15}.

As the upper part of the flux-rope $\Omega$ loop is conveyed upward, we expect it will expand in orthogonal cross-section.  In Figure \ref{fig2}b, we have arbitrarily given the top of the $\Omega$ loop a diameter that is 5/3 bigger than in Figure \ref{fig2}a; the loop's cross-sectional diameter has increased from 30,000 km in Figure \ref{fig2}a to 50,000 km in Figure \ref{fig2}b.  If the field strength in the protrusion in Figure \ref{fig2}a is $\sim$ 1,000 G, then the field strength in the top of the flux-rope $\Omega$ loop in Figure \ref{fig2} is $\sim$ 360 G.  Thus, if the flux-rope $\Omega$ loops that emerge to make the biggest recently-emerged BMRs are made in the manner depicted in Figure \ref{fig2}, and if the field strength of the initial flux tube at the bottom of the convection zone is $\sim 10^3$ G as we have inferred from the strength of the downflows at the bottom of the convection zone in the \cite{mies08} simulation, then we expect the strength of the field in the top of the $\Omega$ loop when it reaches the bottom of the photosphere to be no stronger than several hundred Gauss.

	Our assumption that the diameter of the top of the rising $\Omega$-loop flux rope expanded from its initial diameter at the bottom of the giant cell by only a factor of 5/3 is another bold assumption that needs testing by appropriate numerical simulations.  Because observed BMRs have width of order half the length of the separation distance between the centroids of the two opposite-polarity flux domains, the width of the $\Omega$-loop top that emerges to become a BMR having D $\sim$ 200,000 km should be no more than $\sim$ 100,000 km, $\sim$ twice that drawn in Figure \ref{fig2}b.  We suppose that draining of plasma from the top of the rising $\Omega$ loop down the loop's two legs keeps the width of the top from becoming more than $\sim$ half the $\Omega$ loop's leg-to-leg span, but this needs to be tested by simulations similar to that of \cite{stei12} but for a giant cell.  The \cite{stei12} simulation made a BMR that is roughly twice as long as it is wide.  We have taken the diameter of the $\Omega$ loop at the stage shown in Figure \ref{fig2}b to be 50,000 km, as drawn, because our guess is that the top of the $\Omega$ loop will flatten and spread laterally to become $\sim$ 100,000 km wide during the complex mass-unloading process of emerging through the photosphere \citep[e.g.,][]{cheu10}.  This guess needs testing by simulations of the emergence through the photosphere, for $\Omega$ loops of giant-cell size.

As Figure \ref{fig2}b depicts, we suppose that during the $\sim$ 15 days in which the middle of the flux-rope $\Omega$ loop is carried to the top of the convection zone, each leg of the $\Omega$ loop is swept into the downflow on its side of the giant cell, so that, at middle depths in the convection zone, each leg of the $\Omega$ loop is centered on the axis of the downflow. We also suppose that, at the same time, horizontal flow toward cell center low in the convection zone, having speed $\sim$ 10 m s$^{-1}$, keeps the lower end of each leg of the $\Omega$ loop $\sim$ 30,000 km away from the downflow's axis and toward the center of the giant cell.

Figure \ref{fig2}b also depicts that once the $\Omega$-loop's legs are in the downflows, the counterclockwise rotation in both downflows gives left-handed twist to the magnetic field in the upper part of the $\Omega$ loop, and gives right-handed twist to the field in the lower legs, in the feet, and in the continuation of the $\Omega$-loop's flux tube in the toroidal field band.  We have arbitrarily chosen to depict the field twist that would be given to the $\Omega$-loop flux rope if each of the two downflow swirls were to wind the field about its axis by about one and a half turns during the rise of the $\Omega$ loop.  Thus, the BMR made by the emerged flux-rope $\Omega$ loop will have left-handed overall magnetic twist, in agreement with the observation that most sunspot BMRs in the northern hemisphere have left-handed overall magnetic twist, often seen in coronal images as an inverse-S-shaped sigmoid contortion of the BMR's 3D magnetic field \citep{van15}.

     For the production of a flux-rope $\Omega$ loop by a giant cell, as we have depicted the production in Figure \ref{fig2} and as is thought to be the case in the convection zone \citep{long98}, we assume that the clockwise rotation in the giant cell's upflow is less than the counterclockwise rotation in the downflows at edges of the cell.  So, while some right-handed twist is given to the $\Omega$ loop's magnetic field by the clockwise rotation of the upflow acting on the top part of the rising $\Omega$ loop, that part of the $\Omega$ loop's field is given a greater amount of left-handed twist by the counterclockwise rotations in the two downflows acting on the entrained legs of the $\Omega$ loop.  Consequently, (1) the clockwise rotation of the upflow tilts the plane of the $\Omega$ loop away from east-west so that the west end of the upper part of the $\Omega$ loop is closer to the equator and the east end is farther from the equator (a right-handed twist in the $\Omega$ loop's field), but (2) the tilted upper part of the $\Omega$ loop has overall left-handed axial twist from the downflows in the manner depicted in Figure \ref{fig2}b.  Both that direction of tilt and that sense of axial twist are in accord with what is observed on average in recently-emerged BMR's in the northern hemisphere \citep{van15}.  Thus, Figure \ref{fig2} is a visualization of the so-called $\Sigma$-effect of \cite{long98} that explains why in observed sunspot BMRs in the northern hemisphere the overall bipole tends to have clockwise tilt (which is a right-handed twist in the $\Omega$ loop's magnetic field) but the bipole's magnetic field tends to have net left-handed twist.
     
The drawings in Figures \ref{fig2}a and \ref{fig3}a depict the giant-cell convection flow as though it were laminar.  Simulations of the free convection at all levels of the convection zone \citep{stei89,mies08,stei09} compellingly show that the flow in all convection cells of all sizes -- from granules, through supergranules, to the largest giant cells -- is in reality highly turbulent.  In Figures \ref{fig2} and \ref{fig3}, the schematic depiction of the flow in giant cells purposely ignores the turbulence and shows the flows as laminar to highlight the average direction and spin of the flow in a giant cell.  The simulations show that the convection cells of each size occur chaotically; each crowds out adjacent similar-size extant cells as it is born, grows to its maximum width, and finally is crowded out to extinction by the birth and growth of new cells.  The turbulence and randomness of the free convection results in a substantial minority of the downflows between giant cells having little rotation or rotation direction opposite to that expected from the Sun's global rotation.  That is, the net angular momentum of the inflow to a giant-cell downflow plume is often anticyclonic in direction instead of having the cyclonic direction expected in the plume's hemisphere \citep{mies08}.  If both giant-cell downflows in Figure \ref{fig2}a had anticyclonic direction instead of the depicted cyclonic direction, the resulting twist in the upper part of the flux-rope $\Omega$ loop in Figure \ref{fig2}b would be right-handed instead of left-handed as depicted.  The 3D coronal field of the resulting BMR would have overall S-shaped sigmoid form instead of the usual backward-S shape of recently-emerged sigmoidal BMRs in the northern hemisphere.  From vector magnetograms of BMRs and other signatures of the overall magnetic twist in BMRs, it is observed that a substantial minority ($\sim$ 25\%) of sunspot BMRs have overall magnetic twist opposite in direction to that expected in the BMR's hemisphere \citep{van15}.

\subsection{Smaller Flux-Rope $\Omega$ Loops}

     In our scenario, the flux-rope $\Omega$ loops that become BMRs having flux-centroid-separation distances in the range from about 100,000 km to about 200,000 km are made in the way depicted in Figure \ref{fig2}, by giant cells that span the full depth of the convection zone and have widths equal to the centroid-separation distances of the BMRs.  That is, these flux-rope $\Omega$ loops are made by giant cells ranging in width from about 100,000 km to about 200,000 km.  Because these convection cells reach to the bottom of the convection zone, the BMR flux-rope $\Omega$ loops that they make are made directly from the dynamo-generated toroidal field at the bottom, by the cell ingesting a stitch of that field, as in Figure \ref{fig2}.
     
From SDO/HMI full-disk Doppler velocity observations, \cite{hath15apj} obtained the power spectrum of the Sun's photospheric convection flows.  This power spectrum and simulations of the free convection in the top 2,500 km of the convection zone by \cite{stei89}, in the top 20,000 km by \cite{stei09}, and in the global convection zone at all depths below 15,000 km \citep{mies08} mutually indicate that there is  a continuous size spectrum of continually evolving convection cells, ranging from granules (cell diameter D$_c \sim$ 1,000 km) to the largest giant cells (D$_c \sim$ 200,000 km), the tops of all of which are in the photosphere.  For each size cell, the downflows at the cell edges funnel into the downflows at the edges of larger cells.  As a result, the only downflows at the bottom of the convection zone are downflows at the edges of the largest convection cells, the giant cells wider than about 100,000 km \citep{mies08}.  The funneling of the downflows results in each convection cell that is less than about 100,000 km wide not extending to the bottom of the convection zone but only to a depth comparable to the cell's width.

\begin{figure*}
	\centering
	\includegraphics[width=0.7\linewidth]{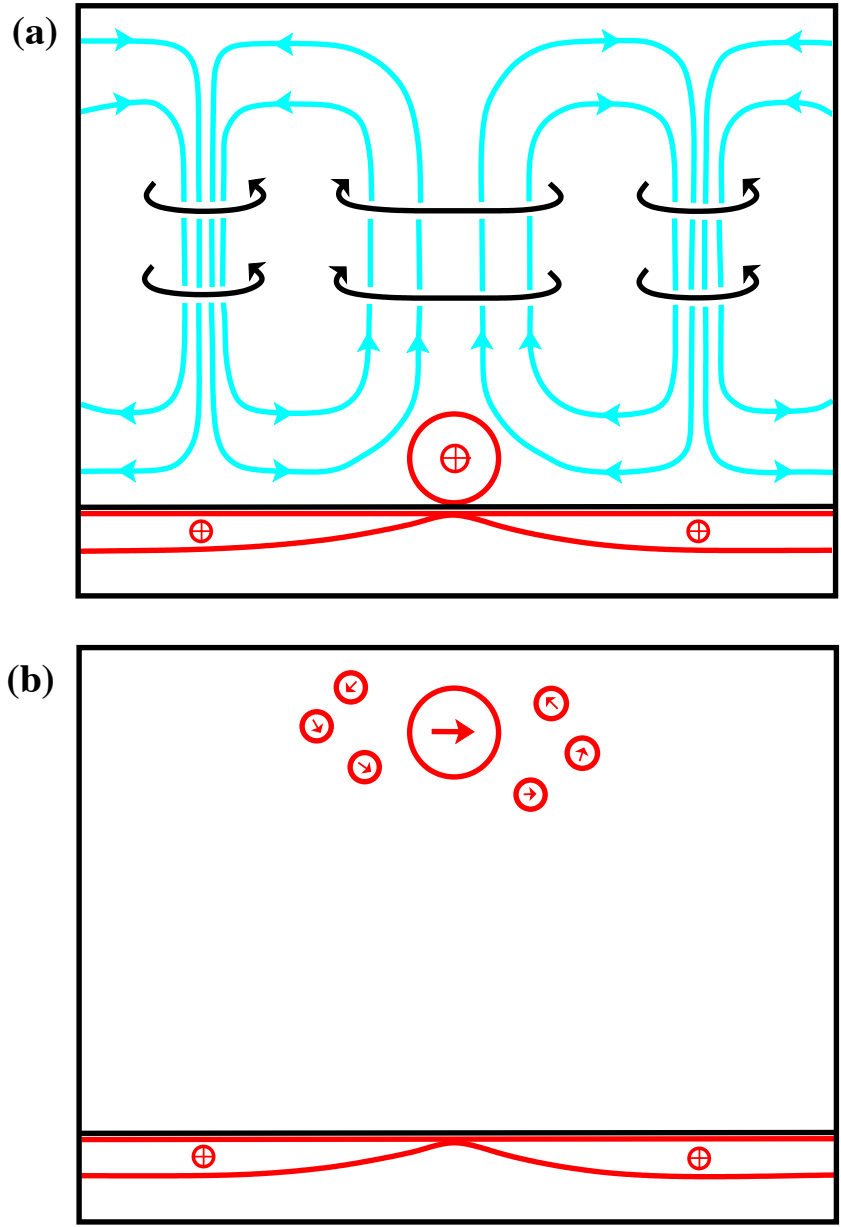}
	\caption{.  Schematic depiction of the production of a failed-BMR $\Omega$ loop by a giant cell from ingested toroidal magnetic field.  The scale, symbolism, and placement $\sim$ 20$\degree$ north of the equator, are the same as in Figure \ref{fig2}, but here the toroidal field and giant cell are viewed from the east, in the direction of the toroidal field, and the vertical cross-section is north-south through the centers of the central giant cell and its two shafts of strong downflow.  In (a), at half a month after the birth of the central giant cell, a flux-tube stitch of the toroidal field has welled up into the bottom of the giant cell and has no twist yet.  In (b), half a month later, because the produced $\Omega$-loop's legs have not been swept into the two strong downflows, the $\Omega$ loop has been given little twist.  That has resulted  in the upper part of the $\Omega$ loop having fragmented into many separate flux tubes that do not directly emerge.  Instead, they remain for about a month in the upper convection zone as horizontal field from which smaller convection cells make BMR $\Omega$ loops of their size.} \label{fig3}
\end{figure*} 

If, as the observations discussed in Section 2 suggest, a BMR of any one size is made by the emergence of a flux-rope $\Omega$ loop of that size and that $\Omega$ loop is made by a convection cell of that size, and if the convection cell makes the $\Omega$ loop in the manner simulated by \cite{stei12} (for a BMR of the size of a small supergranule) and depicted in Figure \ref{fig2} (for a BMR of the size of a big giant cell), then there has to be horizontal magnetic field present at the bottom of the convection cell that the cell can ingest to make into the $\Omega$ loop.  That is, our scenario for BMR $\Omega$-loop production by convection cells requires that at the bottom of each convection cell that makes a BMR flux-rope $\Omega$ loop, there is enough horizontal field for the cell to make the $\Omega$ loop.  In our scenario, the horizontal field for making the $\Omega$ loops for the biggest BMRs, those having centroid-separation distances greater than about 100,000 km, the needed horizontal field is the dynamo-generated toroidal field at the bottom of the convection zone.  For making the $\Omega$ loops for all smaller BMRs, our scenario requires the entire top half of the convection zone be thickly enough populated with horizontal field for the convection cells having their bottoms in the top half of the convection zone to make the flux-rope $\Omega$ loops for all the BMRs smaller than $\sim$ 100,000 km in centroid-separation distance.  Our suggestion for what adequately populates the top half of the convection zone with horizontal field is the leakiness of the turbulent free convection's downward pumping of horizontal field.

The downward pumping of horizontal magnetic field by the convection zone's turbulent free convection is called turbulent pumping or topological pumping \citep[e.g.,][]{tobi01}.  This effect is a prospective agent for pushing horizontal field to the bottom of the convection zone and keeping it there \citep[e.g.,][]{dorc01}.  In particular, turbulent pumping is often advocated for counteracting the buoyancy of the global-dynamo-generated toroidal field at the bottom enough that the dynamo process can build up enough toroidal field flux for a sunspot BMR, before the toroidal field becomes so strong and hence so buoyant that it overcomes the downward pumping and bubbles up in the $\Omega$ loops that make sunspot BMRs \citep[e.g.,][]{dorc01,tobi01,char20}.  Especially in the dynamo scenario of \cite{moor16b}, it is explicitly assumed that all convection-zone horizontal field, either in or not in the dynamo-generated global toroidal field bands, is largely confined to the bottom of the convection zone by the free convection's topological pumping.

Figure \ref{fig2} portrays the downward topological pumping's leakiness in trying to confine the toroidal field band to the bottom of the convection zone.  While the giant-cell downflows do counteract the buoyancy of the toroidal field band and keep it largely held down, the toroidal field leaks upward in the cell centers, where there is no downflow, only upflow.

Figure \ref{fig2} is drawn with the two plumes of strong downflow optimally aligned with the toroidal field for the giant cell to make the flux-rope $\Omega$ loop: the center of the eastern downflow plume is directly east of the center of the western downflow plume.  That is, both downflow plumes sit on the same flux tube of the toroidal field band.  For this alignment, the eastern and western legs of the rising stitch of toroidal field are well positioned to be drawn into the eastern and  western strong downflow plumes, and twisted by them, so that the field stitch is made into a twisted-field flux-rope $\Omega$ loop as in Figure \ref{fig2}b.  If the toroidal field band in each hemisphere is at least $\sim 20\degree$ ($\sim$ 200,000 km) wide in latitude, its 360\degree\ of longitude is always covered by at least a few tens of giant cells.  But no more than a few BMRs the size of giant cells ever occur in either hemisphere per solar rotation.  That is, recently-emerged giant-cell-size BMRs are observed to be few and far between.  Evidently, only a few of the giant cells situated over the toroidal field band make a cell-spanning flux-rope $\Omega$ loop and resulting BMR.  A plausible main reason for this is that a giant cell rarely has two strong downflow plumes aligned directly east-west as in Figure \ref{fig2}.

We conjecture that many of the giant cells that sit on a toroidal field band do not have two roughly equal strong downflow plumes on opposite sides of the cell, and that of those that do have such a pair of downflow plumes, the alignment of the two plumes is usually not nearly east-west as in Figure \ref{fig2}.  For a large majority of giant cells sitting on the toroidal field band and having such a pair of downflow plumes, we suppose that, instead of Figure \ref{fig2}, Figure \ref{fig3} is more representative of the alignment of the pair of downflow plumes with respect the toroidal field.  Like the drawing in Figure \ref{fig2}a, the drawing in Figure \ref{fig3}a is centered on a giant cell that sits $\sim 20\degree$ north of the equator.  But the view direction in Figure \ref{fig3}a is from the east, straight east-west, in the direction of the toroidal field, which, as it does in Figure \ref{fig2}a, points straight west; north is to the left and south is to the right.  In Figure \ref{fig3}a, the alignment of the pair of strong downflow plumes is north-south, orthogonal to the toroidal field.  Figure \ref{fig3}, in the manner of Figure \ref{fig2}, depicts the flow and magnetic field in or centered on a north-south vertical cross-section through the center of the giant cell and the centers of each of the two downflow plumes.  Figure \ref{fig3}a depicts the situation about half a month after the birth of the giant cell.  It is supposed that there are weaker downflows on the east and west sides of the giant cell (in front of and behind the plane of the drawing) that keep the toroidal field band held down there, and that a stitch of the toroidal field band had by now buoyed up in the middle of the bottom of the cell, as shown in Figure \ref{fig3}a.  The red circle in Figure \ref{fig3}a is the cross-section of the summit of the flux-tube stitch, which is being ingested up into the giant cell.

Figure \ref{fig3}b shows the cross-section of the top of the east-west $\Omega$ loop that we envision has been made by the convection flow in the giant cell by about half a month after the time of Figure \ref{fig3}a.  Because the legs of the $\Omega$ loop have not been swept into the two strong downflow plumes, we suppose that the upper half of the $\Omega$ loop is much less twisted than in Figure \ref{fig2}b, and, as a result of having little twist, has not held itself together by its twist to be a single twisted flux rope as in Figure \ref{fig2}b, but has fragmented into a loose bundle of many strands, roughly in the manner of the fragmentation found in simulations of flux ropes of little twist as the flux rope buoyantly rises through the convection zone \citep{fan09}.  Each strand has some fraction of the total magnetic flux in the initial flux-tube stitch of the toroidal field band.  Coriolis-effect clockwise rotation of the upflow has turned the top half of the $\Omega$ loop clockwise (viewed from above).  The legs of the $\Omega$ loop have been swept into the weak downflows on the east and west sides of the giant cell.  The counterclockwise rotation of each of those downflows has given weak left-handed twist to the bundle of flux tubes comprising the top half of the $\Omega$ loop.  The magnetic field direction expected from the two opposite Coriolis-effect twists in combination is shown in Figure \ref{fig3}b by the short arrow in each of the flux-tube cross-sections in the cross-section of the $\Omega$-loop bundle.

We further conjecture that because of having less twist, the strands of each giant-cell-made fragmented $\Omega$ loop are less buoyant than the flux-rope $\Omega$ loop in Figure \ref{fig2}b, not buoyant enough to overcome the downward topological pumping of the free convection enough to emerge through the photosphere and make a BMR that spans the giant cell.  Instead, we envision that the field in the top of each such big fragmented $\Omega$ loop resides in the upper half of the convection zone as roughly horizontal field for $\sim$ 1 month before it is gradually pumped down out of the upper half to the bottom by the downward pumping of the free convection.  Thus, by most giant cells that sit on the toroidal field band making fragmented $\Omega$ loops that fail to directly emerge to become a giant-cell-spanning BMR -- as the flux-rope $\Omega$ loop in Figure \ref{fig2}b is supposed to do -- we envision the upper half of the convection zone to be thickly populated by horizontal field that all smaller convection cells can feed on to make BMRs of their size.

During each 11-year sunspot cycle, sunspot BMRs emerge only in two broad latitudinal bands bracketing the equator, each $\sim$ 20\degree\ wide.  Each band centers about 30\degree\ from the equator at the start of each new 11-year cycle, and steadily drifts equatorward at $\sim$ 1 m s$^{-1}$ to end centered a few degrees from the equator \citep{hath15lrsp}.  Each new band of sunspot BMRs is the continuation of a similar wide band of magnetic activity that begins $\sim$ 10 years earlier at $\sim$ 55\degree\ from the equator, also steadily drifts equatorward at $\sim$ 1 m s$^{-1}$, and starts having sunspot BMRs in it when it reaches $\sim$ 30\degree\ from the equator.  The $\sim$ 20-year progression of the pair of magnetic activity bands -- one in the northern hemisphere and one in the southern hemisphere -- from its start at $\sim \pm$55\degree\ from the equator to the near-equator end of the pair of sunspot-BMR bands that it becomes is called an extended solar activity cycle \citep{wils88}.

At latitudes poleward of $\sim$ 30\degree, the two magnetic activity bands have no sunspot BMRs.  Instead, ephemeral regions, BMRs too small to have sunspots, emerge in each wide band.  The magnetic field of the emerged $\Omega$ loop in these spotless BMRs tends to have the same east-west direction as the sunspot BMRs will have once the band has approached closer than $\sim$ 30\degree\ to the equator \citep{wils88,savc09}.  That is evidence that each hemisphere's global-dynamo-generated toroidal field band -- that becomes the source of that hemisphere's sunspot BMRs in each sunspot cycle -- originates at $\sim$ 55\degree\ from the equator, is the source of the spotless-BMR band that begins with it at $\sim$ 55\degree, and drifts equatorward with it at $\sim$ 1 m s$^{-1}$ \citep{wils88}.  For some reason, the toroidal field band is the source of only spotless small BMRs when it is poleward of $\sim$ 30\degree, but when equatorward of $\sim$ 30\degree, it is the source of sunspot BMRs of all sizes as well as many -- if not all -- spotless smaller BMRs.

In accord with the \cite{moor16b} amended Babcock scenario for the solar-cycle dynamo, we expect the following is the basic reason that no sunspot BMRs emerge from a toroidal field band until the band is within $\sim$ 30\degree\ of the equator.  We expect that, before then, the dynamo process that generates the toroidal field band has not yet given it enough flux for a giant cell to make from a stitch of it an $\Omega$ loop having enough flux for making a sunspot BMR.  For a toroidal field band poleward of $\sim$ 30\degree, we conjecture that, when a giant cell sitting on it makes from a stitch of it -- in the manner of Figures \ref{fig2} and \ref{fig3} -- an $\Omega$ loop of the cell's size, the $\Omega$ loop has too little flux to be buoyant enough to directly emerge as a BMR the size of the giant cell.  We posit that, instead of directly emerging as a giant-cell-size BMR, the top of the giant-cell-made $\Omega$ loop becomes horizontal field residing near the top of the convection zone for about a month before much of it is pumped back down to the bottom of the convection
zone.  We suppose that provides the horizontal field that ephemeral-region-size convection cells can ingest to make flux-rope $\Omega$ loops for the ephemeral-region BMRs in the magnetic activity band when it is poleward of $\sim$ 30\degree.

\cite{gilm18} presents a linearized stability analysis of a model toroidal field layer residing in the tachocline.  The tachocline is the solar-rotation transition layer between the latitudinal differential rotation of the bottom of the convection zone and the solid-body rotation of the convectively-stable solar interior below the convection zone.  The analysis investigates the effect of the radial gradient in rotation speed on the buoyancy instability of the model toroidal field.  Poleward of $\sim \pm$30\degree\ from the equator, rotation in the tachocline is retrograde relative to the interior and the relative rotation speed increases with radial distance up through the tachocline.  Gilman's analysis finds that the negative sense of the radial differential rotation allows the toroidal field layer to be buoyantly unstable for any field strength.  Perhaps that results in the toroidal field layer poleward of $\sim$ 30\degree\ being so leaky against confinement by the giant-cell downflows that the dynamo process cannot build up enough toroidal-field flux to make sunspot BMRs, but only enough to make ephemeral regions \citep{gilm18}.

     Equatorward of $\sim \pm$30\degree, rotation in the tachocline is prograde relative to the interior and the relative rotation speed increases with radial distance up through the tachocline.  Gilman's analysis finds that the positive sense of radial differential rotation and its magnitude render the toroidal field layer buoyantly stable for field strengths up to a latitude-dependent limit.  At the equator, the limit is about 9 x 10$^3$ G and at $\pm$ 20\degree\ latitudes the limit is about 6 x 10$^3$ G.  This suggests that at 20\degree\ the toroidal field strength could be up to three times stronger than our $\sim$ 2 x 10$^3$ G limit obtained from the downward dynamic pressure of the giant-cell downflows in the \cite{mies08} simulation.  A three-times stronger toroidal field in Figure \ref{fig2} would give the $\Omega$ loop $\sim$ 3 x 10$^{22}$ Mx of flux instead of the $\sim$ 1 x 10$^{22}$ Mx that we found from assuming a strength of $\sim$ 2 x 10$^3$ G for the toroidal field band.  Several of the giant-cell-size BMRs measured by \cite{wang89} had about 3 x 10$^{22}$ Mx in the leading-polarity flux domain.
     
Another interesting relevant idea is that before the dynamo process can build the strength of the toroidal field at $\sim$ 20\degree\ to 6 x 10$^3$ G, large-scale dynamics of the tachocline might heave large patches of the toroidal field band up into the bottom of the convection zone \citep{dikp05,dikp20}.  A typical raised patch might span two or three giant cells.  Giant cells plausibly could feed on the raised patches of the toroidal field band to make giant-cell-size $\Omega$ loops and their BMRs.  Perhaps this could be why most BMRs of $\sim$ 200,000 km span have, as the measurements of \cite{wang89} show, leading-polarity flux closer to 1 x 10$^{22}$ Mx than 3 x 10$^{22}$ Mx.

\section{Summary and Discussion}

 This paper envisions a conceptual or heuristic model for how a solar convection cell of any size might make a magnetic-flux-rope $\Omega$ loop that emerges to become a bipolar magnetic region (BMR) of that size.  How the model is based on some well-known BMR observations and solar convection simulations, and how the model works, in summary, are the following.

     \cite{stei12} simulated the production of a BMR of the size of a small supergranule by the emergence of a flux-rope $\Omega$ loop of the BMR's size.  In the simulation, the flux-rope $\Omega$ loop is made by a convection cell of the BMR's size, by the cell ingesting horizontal magnetic field placed at the bottom of the cell. This simulation suggests to us that the magnetic field that emerges to become a BMR of any size -- from as little as granules to as big as giant cells (the convection zone's biggest cells of free convection) -- might be a flux-rope $\Omega$ loop that is made by a convection cell of the BMR's size in the manner of the simulation.  On the basis of the following three observations, the present paper promotes this idea that BMRs of all sizes are emerged $\Omega$ loops, each of which is made by a convection cell and spans that cell.  First, BMRs of all sizes have the magnetic shape and the manner of emergence expected for an emerging flux-rope $\Omega$ loop.  Second, it appears that the $\Omega$ loops that emerge to become the smallest BMRs are each made by the granule convection cell in which it emerges.  Third, the biggest observed recently-emerged BMRs and the biggest giant cells in the \cite{mies08} simulation of the global convection zone both have about the same lateral span and that span is comparable to the depth of the convection zone.  From these observations, we envision how the giant cells that sit on toroidal field generated at the bottom of the convection zone by the global dynamo process might make, from stitches of the toroidal field, $\Omega$ loops that span the giant cell.  Depending on the alignment of the giant cell's downflows with the toroidal field, the generated $\Omega$ loop is one or the other of two kinds.  If the giant cell has a pair of strong downflow plumes on opposite sides, and if the pair of plumes is nearly enough aligned in the direction of the toroidal field, then the generated $\Omega$ loop is a single flux rope and much of the top of the $\Omega$ loop emerges through the photosphere to become a BMR that spans the giant cell.  If the giant cell has no such pair of strong downflow plumes, or if the pair of plumes is not nearly enough aligned with the toroidal field, then the generated $\Omega$ loop is a loose bundle of many separate flux tubes that do not directly emerge as BMRs.  Instead, the dispersed flux tubes of the tops these $\Omega$ loops populate the top half of the convection zone as horizontal magnetic field that smaller convection cells feed on to make flux-rope $\Omega$ loops of their size.  Each of these smaller $\Omega$ loops emerges as a BMR of the size of the convection cell that made it.

As Figure \ref{fig2}b indicates, in the envisioned way of making flux-rope $\Omega$ loops by giant cells ingesting flux-tube stitches from the toroidal field band, the making of each $\Omega$ loop that has left-handed twist in its top part that emerges as a big BMR puts an equal amount of right-handed twist into the $\Omega$ loop's lower legs and their extensions into the toroidal field.  This raises the possibility that, if the global dynamo process generates toroidal field having little or no twist, then, in the northern hemisphere, after enough left-handed-twist flux-rope $\Omega$-loop tops have been made by giant cells, as in Figure \ref{fig2}, say by early in the declining phase of a sunspot cycle, the toroidal field, then and in the remainder of the decline to cycle minimum, might have so much right-handed twist that a majority of large-BMR-making flux-rope $\Omega$-loop tops have net right-handed twist when they emerge. \cite{tiw09b} report possible evidence for this late-in-the-cycle hemispheric-twist-rule violation anticipated from our picture of the generation of flux-rope $\Omega$ loops by giant cells.  From high-resolution vector magnetograms they measured the global (net) twist in the magnetic field in each of 47 sunspots in the declining phase of sunspot Cycle 23.  For 42 of the 47 sunspots, the direction of the overall net twist in the sunspot's magnetic field was the opposite of the direction expected for the sunspot's hemisphere. \cite{hao11} also found results consistent with those of \cite{tiw09b}.

For the global solar dynamo process, \cite{moor16b} present a scenario that is an extension of the original \cite{babc61} scenario for the dynamo process that sustains the 11-year sunspot cycle.  To a limited degree, the \cite{moor16b} scenario is demonstrated by the solar dynamo simulation of \cite{yeat13}.  The \cite{moor16b}  scenario envisions how the polar field that is present at the onset of each new sunspot cycle might be made a la \cite{babc61} from the BMRs of the previous cycle (called the old cycle) in a way such that, at about two years after the onset of the old sunspot cycle, the new-cycle toroidal field band starts being generated at latitudes above those of the old-cycle toroidal field band, in agreement with the observed so-called extended solar activity cycle \citep{wils88}.

A key requirement of the \cite{moor16b} dynamo scenario is that the convection zone's turbulent free convection, via turbulent pumping, pushes horizontal field to the bottom and holds it there.  It is supposed that the downward turbulent pumping is what makes the toroidal field reside at the bottom of the convection zone as the toroidal field is built there from horizontal poloidal field by latitudinal differential rotation.  From assuming that the strength of the toroidal field for large-sunspot-BMR production is set by the giant-cell downflows at the bottom of the convection zone, which the \cite{mies08} simulation finds to be $\sim$ 10 m s$^{-1}$, the strength of the toroidal field in our Figure \ref{fig2} scenario for giant-cell $\Omega$-loop production is $\sim 2 \times 10^3$ G.  With that field strength and the $\sim$ 200,000 km width of the biggest giant cells in the \cite{mies08} simulation, our Figure \ref{fig2} scenario plausibly explains why the magnetic flux in each $\Omega$ loop for the biggest recently-emerged BMRs (which have their two opposite-polarity flux centroids separated by $\sim$ 200,000 km) is $\sim 10^{22}$ Mx.  In this way, our scenario for making the biggest BMR $\Omega$ loops from giant cells and the \cite{mies08} simulation of the global convection zone together (1) support the \cite{moor16b} amended Babcock scenario for the sunspot-cycle dynamo process, and (2) question the widely-held idea that sunspot-BMR $\Omega$ loops are buoyant flux ropes in which the field strength is $\sim 10^5$ G when the flux rope begins to rise from the bottom of the convection zone.

The \cite{mies08} global simulation of the convection zone supports the \cite{moor16b} dynamo scenario in another way as well.  The \cite{moor16b} scenario supposes that, to match the $\sim$ 1 m s$^{-1}$ equatorward drift of each magnetic activity band of the extended solar cycle, the toroidal field band that is the source of the active magnetic field and sits under it at the bottom of the convection zone drifts equatorward at $\sim$ 1 m s$^{-1}$, conveyed at that speed by equatorward meridional flow.  In the \cite{mies08} simulation, at the bottom of the convection zone there is equatorward meridional flow of that magnitude.

In Figures \ref{fig2} and \ref{fig3}, the top of the toroidal field band generated by the global dynamo process is arbitrarily placed at the middle of the tachocline.  The proposed scenario for making cell-spanning $\Omega$ loops by giant cells is practically the same if the toroidal field band is placed 15,000 km higher so that its bottom is at the middle of the tachocline.  This higher placement of the toroidal field band is perhaps preferable to the lower placement depicted in Figures \ref{fig2} and \ref{fig3} if, as in the sunspot-cycle dynamo scenario of \cite{moor16b}, the toroidal field for the sunspot cycle is at the bottom of the convection zone, is generated by latitudinal differential rotation, and is swept equatorward at $\sim$ 1 m s$^{-1}$ by meridional flow.

Whether giant cells can make BMR and failed-BMR $\Omega$ loops of their size from underlying toroidal field in the manner that we have schematically envisioned here remains to be tested by simulations that are similar to that of \cite{stei12}, but span the full depth of the convection zone instead of only the top 20,000 km.  Any such simulation is beyond the scope of the present paper.  We hope that our scenario for making BMR $\Omega$ loops of all sizes will lead to simulations that can test it.

\acknowledgments
 Insightful and helpful comments from the referee improved the structure, clarity, and strength of the paper.  RLM, NKP, and ACS were supported by NASA's Heliophysics Guest Investigators Program.  SKT gratefully acknowledges support by NASA contract NNM07AA01C (Hinode).  NKP also acknowledges support from NASA's SDO/AIA (NNG04EA00C). Hinode is a Japanese mission developed and launched
by ISAS/JAXA, collaborating with NAOJ as a domestic partner, NASA and STFC (UK) as international partners. Scientific operation of the Hinode mission is conducted by the Hinode science team organized at ISAS/JAXA. This team mainly consists of scientists from institutes in the partner countries. Support for the post-launch operation is provided by JAXA and NAOJ (Japan), STFC (UK), NASA, ESA, and NSC (Norway).  The HMI data are courtesy of NASA/SDO and the HMI science team. This work has made use of NASA ADSABS and Solar Software.


\begin{thebibliography}{}
	\expandafter\ifx\csname natexlab\endcsname\relax\def\natexlab#1{#1}\fi
	\providecommand{\url}[1]{\href{#1}{#1}}
	\providecommand{\dodoi}[1]{doi:~\href{http://doi.org/#1}{\nolinkurl{#1}}}
	\providecommand{\doeprint}[1]{\href{http://ascl.net/#1}{\nolinkurl{http://ascl.net/#1}}}
	\providecommand{\doarXiv}[1]{\href{https://arxiv.org/abs/#1}{\nolinkurl{https://arxiv.org/abs/#1}}}
	
	\bibitem[{{Babcock}(1961)}]{babc61}
	{Babcock}, H.~W. 1961, \apj, 133, 572, \dodoi{10.1086/147060}
	
	\bibitem[{{Bruzek}(1977)}]{bruz77}
	{Bruzek}, A. 1977, Astrophysics and Space Science Library, Vol.~69, {Spots and
		Faculae}, ed. A.~{Bruzek} \& C.~J. {Durrant}, 71
	
	\bibitem[{{Charbonneau}(2020)}]{char20}
	{Charbonneau}, P. 2020, Living Reviews in Solar Physics, 17, 4,
	\dodoi{10.1007/s41116-020-00025-6}
	
	\bibitem[{{Cheung} {et~al.}(2010){Cheung}, {Rempel}, {Title}, \&
		{Sch{\"u}ssler}}]{cheu10}
	{Cheung}, M.~C.~M., {Rempel}, M., {Title}, A.~M., \& {Sch{\"u}ssler}, M. 2010,
	\apj, 720, 233, \dodoi{10.1088/0004-637X/720/1/233}
	
	\bibitem[{{Dikpati} \& {Gilman}(2005)}]{dikp05}
	{Dikpati}, M., \& {Gilman}, P.~A. 2005, \apjl, 635, L193,
	\dodoi{10.1086/499626}
	
	\bibitem[{{Dikpati} \& {McIntosh}(2020)}]{dikp20}
	{Dikpati}, M., \& {McIntosh}, S.~W. 2020, Space Weather, 18, e02109,
	\dodoi{10.1029/2019SW002109}
	
	\bibitem[{{Dorch} \& {Nordlund}(2001)}]{dorc01}
	{Dorch}, S.~B.~F., \& {Nordlund}, {\r{A}}. 2001, \aap, 365, 562,
	\dodoi{10.1051/0004-6361:20000141}
	
	\bibitem[{{Fan}(2009)}]{fan09}
	{Fan}, Y. 2009, Living Reviews in Solar Physics, 6, 4,
	\dodoi{10.12942/lrsp-2009-4}
	
	\bibitem[{{Gilman}(2018)}]{gilm18}
	{Gilman}, P.~A. 2018, \apj, 853, 65, \dodoi{10.3847/1538-4357/aaa4f4}
	
	\bibitem[{{Guenther} {et~al.}(1992){Guenther}, {Demarque}, {Kim}, \&
		{Pinsonneault}}]{guen92}
	{Guenther}, D.~B., {Demarque}, P., {Kim}, Y.~C., \& {Pinsonneault}, M.~H. 1992,
	\apj, 387, 372, \dodoi{10.1086/171090}
	
	\bibitem[{{Hale} \& {Nicholson}(1925)}]{haln25}
	{Hale}, G.~E., \& {Nicholson}, S.~B. 1925, \apj, 62, 270,
	\dodoi{10.1086/142933}
	
	\bibitem[{{Hao} \& {Zhang}(2011)}]{hao11}
	{Hao}, J., \& {Zhang}, M. 2011, \apjl, 733, L27,
	\dodoi{10.1088/2041-8205/733/2/L27}
	
	\bibitem[{{Hathaway}(2015)}]{hath15lrsp}
	{Hathaway}, D.~H. 2015, Living Reviews in Solar Physics, 12, 4,
	\dodoi{10.1007/lrsp-2015-4}
	
	\bibitem[{{Hathaway} {et~al.}(2015){Hathaway}, {Teil}, {Norton}, \&
		{Kitiashvili}}]{hath15apj}
	{Hathaway}, D.~H., {Teil}, T., {Norton}, A.~A., \& {Kitiashvili}, I. 2015,
	\apj, 811, 105, \dodoi{10.1088/0004-637X/811/2/105}
	
	\bibitem[{{Hathaway} {et~al.}(2013){Hathaway}, {Upton}, \&
		{Colegrove}}]{hath13}
	{Hathaway}, D.~H., {Upton}, L., \& {Colegrove}, O. 2013, Science, 342, 1217,
	\dodoi{10.1126/science.1244682}
	
	\bibitem[{{Ichimoto} {et~al.}(2008){Ichimoto}, {Lites}, {Elmore}, {Suematsu},
		{Tsuneta}, {Katsukawa}, {Shimizu}, {Shine}, {Tarbell}, {Title}, {Kiyohara},
		{Shinoda}, {Card}, {Lecinski}, {Streander}, {Nakagiri}, {Miyashita},
		{Noguchi}, {Hoffmann}, \& {Cruz}}]{ichi08}
	{Ichimoto}, K., {Lites}, B., {Elmore}, D., {et~al.} 2008, \solphys, 249, 233,
	\dodoi{10.1007/s11207-008-9169-9}
	
	\bibitem[{{Innes} \& {Teriaca}(2013)}]{inne13}
	{Innes}, D.~E., \& {Teriaca}, L. 2013, \solphys, 282, 453,
	\dodoi{10.1007/s11207-012-0199-y}
	
	\bibitem[{{Ishikawa} \& {Tsuneta}(2010)}]{ishit10}
	{Ishikawa}, R., \& {Tsuneta}, S. 2010, \apjl, 718, L171,
	\dodoi{10.1088/2041-8205/718/2/L171}
	
	\bibitem[{{Ishikawa} {et~al.}(2010){Ishikawa}, {Tsuneta}, \&
		{Jur{\v{c}}{\'a}k}}]{ishi10}
	{Ishikawa}, R., {Tsuneta}, S., \& {Jur{\v{c}}{\'a}k}, J. 2010, \apj, 713, 1310,
	\dodoi{10.1088/0004-637X/713/2/1310}
	
	\bibitem[{{Ishikawa} {et~al.}(2008){Ishikawa}, {Tsuneta}, {Ichimoto}, {Isobe},
		{Katsukawa}, {Lites}, {Nagata}, {Shimizu}, {Shine}, {Suematsu}, {Tarbell}, \&
		{Title}}]{ishi08}
	{Ishikawa}, R., {Tsuneta}, S., {Ichimoto}, K., {et~al.} 2008, \aap, 481, L25,
	\dodoi{10.1051/0004-6361:20079022}
	
	\bibitem[{{Kosugi} {et~al.}(2007){Kosugi}, {Matsuzaki}, {Sakao}, {Shimizu},
		{Sone}, {Tachikawa}, {Hashimoto}, {Minesugi}, {Ohnishi}, {Yamada}, {Tsuneta},
		{Hara}, {Ichimoto}, {Suematsu}, {Shimojo}, {Watanabe}, {Shimada}, {Davis},
		{Hill}, {Owens}, {Title}, {Culhane}, {Harra}, {Doschek}, \& {Golub}}]{kosu07}
	{Kosugi}, T., {Matsuzaki}, K., {Sakao}, T., {et~al.} 2007, \solphys, 243, 3,
	\dodoi{10.1007/s11207-007-9014-6}
	
	\bibitem[{{Lites} {et~al.}(2008){Lites}, {Kubo}, {Socas-Navarro}, {Berger},
		{Frank}, {Shine}, {Tarbell}, {Title}, {Ichimoto}, {Katsukawa}, {Tsuneta},
		{Suematsu}, {Shimizu}, \& {Nagata}}]{lites08}
	{Lites}, B.~W., {Kubo}, M., {Socas-Navarro}, H., {et~al.} 2008, \apj, 672,
	1237, \dodoi{10.1086/522922}
	
	\bibitem[{{Longcope} {et~al.}(1998){Longcope}, {Fisher}, \& {Pevtsov}}]{long98}
	{Longcope}, D.~W., {Fisher}, G.~H., \& {Pevtsov}, A.~A. 1998, \apj, 507, 417,
	\dodoi{10.1086/306312}
	
	\bibitem[{{Martres} \& {Bruzek}(1977)}]{mart77}
	{Martres}, M.~J., \& {Bruzek}, A. 1977, Astrophysics and Space Science Library,
	Vol.~69, {Active Regions}, ed. A.~{Bruzek} \& C.~J. {Durrant}, 53
	
	\bibitem[{{Miesch} {et~al.}(2008){Miesch}, {Brun}, {DeRosa}, \&
		{Toomre}}]{mies08}
	{Miesch}, M.~S., {Brun}, A.~S., {DeRosa}, M.~L., \& {Toomre}, J. 2008, \apj,
	673, 557, \dodoi{10.1086/523838}
	
	\bibitem[{{Moore} {et~al.}(2000){Moore}, {Hathaway}, \& {Reichmann}}]{moor00}
	{Moore}, R., {Hathaway}, D., \& {Reichmann}, E. 2000, in AAS/Solar Physics
	Division Meeting, Vol.~31, AAS/Solar Physics Division Meeting \#31, 04.03
	
	\bibitem[{{Moore} \& {Rabin}(1985)}]{moor85}
	{Moore}, R., \& {Rabin}, D. 1985, \araa, 23, 239,
	\dodoi{10.1146/annurev.aa.23.090185.001323}
	
	\bibitem[{Moore {et~al.}(2016)Moore, Cirtain, \& Sterling}]{moor16b}
	Moore, R.~L., Cirtain, J.~W., \& Sterling, A.~C. 2016, Babcock Redux: An
	Ammendment of Babcock's Schematic of the Sun's Magnetic Cycle.
	\newblock \doarXiv{1606.05371}
	
	\bibitem[{{Moore} {et~al.}(2011){Moore}, {Sterling}, {Cirtain}, \&
		{Falconer}}]{moor11}
	{Moore}, R.~L., {Sterling}, A.~C., {Cirtain}, J.~W., \& {Falconer}, D.~A. 2011,
	\apjl, 731, L18, \dodoi{10.1088/2041-8205/731/1/L18}
	
	\bibitem[{{Panesar} {et~al.}(2019){Panesar}, {Sterling}, {Moore}, {Winebarger},
		{Tiwari}, {Savage}, {Golub}, {Rachmeler}, {Kobayashi}, {Brooks}, {Cirtain},
		{De Pontieu}, {McKenzie}, {Morton}, {Peter}, {Testa}, {Walsh}, \&
		{Warren}}]{pane19}
	{Panesar}, N.~K., {Sterling}, A.~C., {Moore}, R.~L., {et~al.} 2019, \apjl, 887,
	L8, \dodoi{10.3847/2041-8213/ab594a}
	
	\bibitem[{{Parker}(1955)}]{parker55}
	{Parker}, E.~N. 1955, \apj, 121, 491, \dodoi{10.1086/146010}
	
	\bibitem[{{Parker}(1975)}]{parker75}
	---. 1975, \apj, 201, 494, \dodoi{10.1086/153911}
	
	\bibitem[{{Raouafi} {et~al.}(2010){Raouafi}, {Georgoulis}, {Rust}, \&
		{Bernasconi}}]{raou10}
	{Raouafi}, N.~E., {Georgoulis}, M.~K., {Rust}, D.~M., \& {Bernasconi}, P.~N.
	2010, \apj, 718, 981, \dodoi{10.1088/0004-637X/718/2/981}
	
	\bibitem[{{Savcheva} {et~al.}(2009){Savcheva}, {Cirtain}, {DeLuca}, \&
		{Golub}}]{savc09}
	{Savcheva}, A., {Cirtain}, J.~W., {DeLuca}, E.~E., \& {Golub}, L. 2009, \apjl,
	702, L32, \dodoi{10.1088/0004-637X/702/1/L32}
	
	\bibitem[{{Shibata} {et~al.}(2007){Shibata}, {Nakamura}, {Matsumoto}, {Otsuji},
		{Okamoto}, {Nishizuka}, {Kawate}, {Watanabe}, {Nagata}, {UeNo}, {Kitai},
		{Nozawa}, {Tsuneta}, {Suematsu}, {Ichimoto}, {Shimizu}, {Katsukawa},
		{Tarbell}, {Berger}, {Lites}, {Shine}, \& {Title}}]{shib07}
	{Shibata}, K., {Nakamura}, T., {Matsumoto}, T., {et~al.} 2007, Science, 318,
	1591, \dodoi{10.1126/science.1146708}
	
	\bibitem[{{Simon} \& {Weiss}(1968)}]{simo68}
	{Simon}, G.~W., \& {Weiss}, N.~O. 1968, \zap, 69, 435
	
	\bibitem[{{Spruit}(2011)}]{spru11}
	{Spruit}, H.~C. 2011, {Theories of the Solar Cycle: A Critical View}, ed. M.~P.
	{Miralles} \& J.~{S{\'a}nchez Almeida}, Vol.~4, 39
	
	\bibitem[{{Stein} \& {Nordlund}(1989)}]{stei89}
	{Stein}, R.~F., \& {Nordlund}, A. 1989, \apjl, 342, L95, \dodoi{10.1086/185493}
	
	\bibitem[{{Stein} \& {Nordlund}(2012)}]{stei12}
	{Stein}, R.~F., \& {Nordlund}, {\r{A}}. 2012, \apjl, 753, L13,
	\dodoi{10.1088/2041-8205/753/1/L13}
	
	\bibitem[{{Stein} {et~al.}(2009){Stein}, {Nordlund}, {Georgoviani}, {Benson},
		\& {Schaffenberger}}]{stei09}
	{Stein}, R.~F., {Nordlund}, {\r{A}}., {Georgoviani}, D., {Benson}, D., \&
	{Schaffenberger}, W. 2009, in Astronomical Society of the Pacific Conference
	Series, Vol. 416, Solar-Stellar Dynamos as Revealed by Helio- and
	Asteroseismology: GONG 2008/SOHO 21, ed. M.~{Dikpati}, T.~{Arentoft},
	I.~{Gonz{\'a}lez Hern{\'a}ndez}, C.~{Lindsey}, \& F.~{Hill}, 421
	
	\bibitem[{{Sterling} \& {Moore}(2020)}]{ster20}
	{Sterling}, A.~C., \& {Moore}, R.~L. 2020, \apjl, 896, L18,
	\dodoi{10.3847/2041-8213/ab96be}
	
	\bibitem[{{Suematsu} {et~al.}(2008){Suematsu}, {Tsuneta}, {Ichimoto},
		{Shimizu}, {Otsubo}, {Katsukawa}, {Nakagiri}, {Noguchi}, {Tamura}, {Kato},
		{Hara}, {Kubo}, {Mikami}, {Saito}, {Matsushita}, {Kawaguchi}, {Nakaoji},
		{Nagae}, {Shimada}, {Takeyama}, \& {Yamamuro}}]{suem08}
	{Suematsu}, Y., {Tsuneta}, S., {Ichimoto}, K., {et~al.} 2008, \solphys, 249,
	197, \dodoi{10.1007/s11207-008-9129-4}
	
	\bibitem[{{Thompson} {et~al.}(2003){Thompson}, {Christensen-Dalsgaard},
		{Miesch}, \& {Toomre}}]{thom03}
	{Thompson}, M.~J., {Christensen-Dalsgaard}, J., {Miesch}, M.~S., \& {Toomre},
	J. 2003, \araa, 41, 599, \dodoi{10.1146/annurev.astro.41.011802.094848}
	
	\bibitem[{{Title} {et~al.}(1989){Title}, {Tarbell}, {Topka}, {Ferguson},
		{Shine}, \& {SOUP Team}}]{titl89}
	{Title}, A.~M., {Tarbell}, T.~D., {Topka}, K.~P., {et~al.} 1989, \apj, 336,
	475, \dodoi{10.1086/167026}
	
	\bibitem[{{Tiwari} {et~al.}(2009){Tiwari}, {Venkatakrishnan}, \&
		{Sankarasubramanian}}]{tiw09b}
	{Tiwari}, S.~K., {Venkatakrishnan}, P., \& {Sankarasubramanian}, K. 2009,
	\apjl, 702, L133, \dodoi{10.1088/0004-637X/702/2/L133}
	
	\bibitem[{{Tiwari} {et~al.}(2019){Tiwari}, {Panesar}, {Moore}, {De Pontieu},
		{Winebarger}, {Golub}, {Savage}, {Rachmeler}, {Kobayashi}, {Testa}, {Warren},
		{Brooks}, {Cirtain}, {McKenzie}, {Morton}, {Peter}, \& {Walsh}}]{tiw19}
	{Tiwari}, S.~K., {Panesar}, N.~K., {Moore}, R.~L., {et~al.} 2019, \apj, 887,
	56, \dodoi{10.3847/1538-4357/ab54c1}
	
	\bibitem[{{Tobias} {et~al.}(2001){Tobias}, {Brummell}, {Clune}, \&
		{Toomre}}]{tobi01}
	{Tobias}, S.~M., {Brummell}, N.~H., {Clune}, T.~L., \& {Toomre}, J. 2001, \apj,
	549, 1183, \dodoi{10.1086/319448}
	
	\bibitem[{{Tsuneta} {et~al.}(2008){Tsuneta}, {Ichimoto}, {Katsukawa}, {Nagata},
		{Otsubo}, {Shimizu}, {Suematsu}, {Nakagiri}, {Noguchi}, {Tarbell}, {Title},
		{Shine}, {Rosenberg}, {Hoffmann}, {Jurcevich}, {Kushner}, {Levay}, {Lites},
		{Elmore}, {Matsushita}, {Kawaguchi}, {Saito}, {Mikami}, {Hill}, \&
		{Owens}}]{tsun08}
	{Tsuneta}, S., {Ichimoto}, K., {Katsukawa}, Y., {et~al.} 2008, \solphys, 249,
	167, \dodoi{10.1007/s11207-008-9174-z}
	
	\bibitem[{{Vaiana} \& {Rosner}(1978)}]{vaia78}
	{Vaiana}, G.~S., \& {Rosner}, R. 1978, \araa, 16, 393,
	\dodoi{10.1146/annurev.aa.16.090178.002141}
	
	\bibitem[{{van Driel-Gesztelyi} \& {Green}(2015)}]{van15}
	{van Driel-Gesztelyi}, L., \& {Green}, L.~M. 2015, Living Reviews in Solar
	Physics, 12, 1, \dodoi{10.1007/lrsp-2015-1}
	
	\bibitem[{{Wang} \& {Sheeley}(1989)}]{wang89}
	{Wang}, Y.~M., \& {Sheeley}, N.~R., J. 1989, \solphys, 124, 81,
	\dodoi{10.1007/BF00146521}
	
	\bibitem[{{Wilson} {et~al.}(1988){Wilson}, {Altrocki}, {Harvey}, {Martin}, \&
		{Snodgrass}}]{wils88}
	{Wilson}, P.~R., {Altrocki}, R.~C., {Harvey}, K.~L., {Martin}, S.~F., \&
	{Snodgrass}, H.~B. 1988, \nat, 333, 748, \dodoi{10.1038/333748a0}
	
	\bibitem[{{Withbroe} \& {Noyes}(1977)}]{with77}
	{Withbroe}, G.~L., \& {Noyes}, R.~W. 1977, \araa, 15, 363,
	\dodoi{10.1146/annurev.aa.15.090177.002051}
	
	\bibitem[{{Yeates} \& {Mu{\~n}oz-Jaramillo}(2013)}]{yeat13}
	{Yeates}, A.~R., \& {Mu{\~n}oz-Jaramillo}, A. 2013, \mnras, 436, 3366,
	\dodoi{10.1093/mnras/stt1818}
	
	\bibitem[{{Zwaan}(1987)}]{zwaa87}
	{Zwaan}, C. 1987, \araa, 25, 83, \dodoi{10.1146/annurev.aa.25.090187.000503}
	
\end{thebibliography}

\end{document}